\documentclass[12pt,a4paper]{article}
\usepackage[latin9]{inputenc}
\usepackage{times}
\usepackage{a4wide}
\usepackage{amssymb}
\usepackage{amsmath}
\usepackage{bbm}
\usepackage{dsfont} 
\usepackage{ulem} 
\usepackage{pifont} 
\usepackage{ifpdf}
\ifpdf
\usepackage[pdftex]{graphicx}
\usepackage[pdftex,unicode,implicit]{hyperref}
\hypersetup{%
  pdftitle    = {All the timelike supersymmetric solutions of all ungauged d=4  supergravities},
  pdfkeywords = {supergravity, supersymmetry, BPS},
  pdfauthor   = {P. Meessen, T. Ortin and Silvia Vaula},
  pdfcreator  = {pdf\LaTeXe\ with package \flqq hyperref\frqq},
  pdfproducer = {pdf\LaTeXe\ with package \flqq hyperref\frqq},
  pdfpagemode = None,  
  pdffitwindow= true,  
  unicode     = true,
  plainpages  = false,
  colorlinks  = true,  
  citecolor   = blue,
  urlcolor    = red,
  linkcolor   = black
}
\newcommand{\hepth}[1]{{\tt \href{http://www.arXiv.org/abs/hep-th/#1}{hep-th/#1}}}

\newcommand{\arxiv}[1]{{\tt \href{http://www.arXiv.org/abs/#1}{arXiv:#1}}}
\else
  \usepackage[dvips]{graphicx}
  \usepackage[unicode,implicit]{hyperref}
  \newcommand{\hepth}[1]{{\tt hep-th/#1}}

  \newcommand{\arxiv}[1]{{\tt arXiv:#1}}
\fi
\makeatletter
\@addtoreset{equation}{section}
\makeatother

\begin{document}

\begin{flushright}
\small
IFT-UAM/CSIC-10-52\\
Oct $7^{\rm th}$, $2010$
\normalsize
\end{flushright}
\begin{center}
\vspace{.5cm}
{\LARGE {\bf Supersymmetric solutions }}\\[.5cm]
{\LARGE {\bf of 4-dimensional supergravities}} 
\vspace{1.2cm}

{\sl\large Tom{\'a}s Ort\'{\i}n}
\footnote{E-mail: {\tt Tomas.Ortin@cern.ch}}

\vspace{.8cm}

{\it Instituto de F\'{\i}sica Te\'orica UAM/CSIC\\
Facultad de Ciencias C-XVI,  C.U.~Cantoblanco,  E-28049 Madrid, Spain}\\

\vspace{3cm}


{\bf Abstract}

\end{center}

\begin{quotation}\small
  We review general and recent results on the characterization and
  construction of timelike supersymmetric solutions of 4-dimensional
  supergravity theories.
\end{quotation}

\vspace{1.5cm} 
\textit{Contribution to the Proceedings of the IV Mexican
  Meeting in Mathematical and Experimental Physics held at El Colegio
  Nacional, México DF.    July 19th to 23rd, 2010} 
\vspace{1.5cm}

\newpage

\pagestyle{plain}


\section{Introduction:  r\^ole and importance of supersymmetric solutions}

Supersymmetry arose almost 40 years ago as a possible symmetry of Nature that
would unify the two seemingly different types of elementary constituents of
the Universe: matter and interactions (\textit{i.e.}~bosons $\phi^{b}$ and
fermions $\phi^{f}$). Much of the interest in supersymmetry is due to the fact
that its presence is crucial for the consistency of (super-) string theories
(and gravity/gauge (AdS/CFT) correspondences) on different vacua.

The local (\textit{gauge}) generalization of supersymmetry, aptly named
\textit{supergravity} requires/implies the coupling of all the fields to
standard (General Relativity (GR)) gravity which is, then, included in the
unification. From a more pedestrian point of view supergravity theories can be
seen as nothing but extensions of GR consisting in a number of fermionic and
bosonic fields coupled to gravity. Many purely bosonic theories can be
supersymmetrized (or \textit{embedded} in a supersymmetric theory) by the simple
addition of fermionic fields in appropriate numbers and species and with
appropriate couplings. The cosmological Einstein-Maxwell theory, for instance,
can be embedded in $N=1,d=4$ supergravity in a number of ways
\cite{Meessen:2010ph}.

At low energies, superstring theories can be effectively described by
supergravity theories\footnote{The reverse is not always true.}. This leads to
an extremely rich interplay between superstring and supergravity theories
which has allowed, for instance, to prove the UV finiteness of $N=8$
supergravity to the fourth loop order, raising the possibility that it may be
a finite quantum field theory of gravity \cite{Bern:2009kd}.

In this talk we are interested in purely bosonic ($\phi^{f} =0$) solutions of
the classical equations of motion of supergravity theories. Since $\phi^{f}=0$
is always a consistent truncation, the solutions of the truncated supergravity
theory (GR coupled to some bosonic fields) are automatically solutions of the
full supergravity theory. Thus, for instance, all the standard solutions of
the cosmological Einstein-Maxwell theory are also purely bosonic solutions of
$N=1,d=4$ supergravity and \textit{vice versa}. These solutions are also
important from the superstring theory point of view: the theory can only be
quantized consistently in backgrounds (\textit{vacua}) which are solutions or
their associated supergravity description\footnote{We would like to stress
  that the proliferation of possible string theory vacua, the so-called
  \textit{landscape problem}, is actually common to all theories containing
  GR.}. Furthermore, the \textit{supersymmetric} solutions, to be defined later,
can also be interpreted as the long-range fields generated by a source which
is a state of the superstring theory. The identification of the sources of
\textit{supersymmetric} (a.k.a.~BPS) black holes in terms of states of
superstring theory on a suitable background is the keystone of the
microscopic interpretation (via the ``gauge dual'') of these black hole's
entropy.

The (unbroken) supersymmetry of the classical solution plays a crucial role in
this and many other problems. This is what makes supersymmetric (BPS)
solutions interesting. Many interesting and well-known GR solutions
(Minkowski, (Anti-) De Sitter, extreme Reissner-Nordstr\"om, $pp$-waves...)
\textit{are} supersymmetric. Let us define this property: a bosonic field
configuration of a supergravity theory (no necessarily solving its equations
of motion) is \textit{supersymmetric} if it is invariant under some
supersymmetry transformations. The transformations generated by the spinor
$\epsilon^{\alpha}(x)$ take the generic form
\begin{equation}
  \delta_{\epsilon} \phi^{b}
 \sim \bar{\epsilon} \phi^{f}\, ,
\hspace{1cm}
  \delta_{\epsilon} \phi^{f}
 \sim \partial \epsilon 
+(\phi^{b} +\bar\phi^{f}\phi^{f}) \epsilon\, ,
\end{equation}
Then, a bosonic configuration is invariant under the transformation generated
by $\epsilon^{\alpha}(x)$ if it satisfies the \textit{Killing Spinor Equations
  (KSEs)}
\begin{equation}
  \delta_{\epsilon} \phi^{f}
\sim \partial \epsilon 
+\phi^{b}\epsilon=0\, .
\end{equation}

This is a generalization of the concept of isometry, an infinitesimal
g.c.t.~generated by $\xi^{\mu}(x)$ that leaves the metric $g_{\mu\nu}$
invariant because it satisfies the Killing Vector Equation.  Each isometry is
associated to a bosonic generator of a (Lie) symmetry algebra
\begin{equation}
\xi_{(I)}^{\mu}(x)\rightarrow P_{I}\, ,
\hspace{1cm}
[P_{I},P_{J}]  =f_{IJ}{}^{K}P_{K}\, .
\end{equation}
Correspondingly, each supersymmetry is associated to an odd generator
of a  (Lie) symmetry superalgebra
\begin{equation}
\epsilon_{(n)}^{\alpha}(x)\rightarrow \mathcal{Q}_{n}\, , 
\hspace{1cm}
[\mathcal{Q}_{n},P_{I}]  =f_{nI}{}^{m} \mathcal{Q}_{m}  \, ,
\hspace{1cm}
\{\mathcal{Q}_{n},\mathcal{Q}_{m}\}  = f_{nm}{}^{I}  P_{I}\, .  
\end{equation}

Every supersymmetric field configuration has a supersymmetry superalgebra. For
instance, the superalgebra of Minkowski spacetime is the Poincar\'e
superalgebra with
\begin{equation}
\{\mathcal{Q}_{\alpha},\mathcal{Q}_{\beta}\}
  = 
(\gamma^{\mu} \mathcal{C})_{\alpha\beta} P_{\mu}\, .  
\end{equation}

The supersymmetric solutions have a number of interesting properties:
\begin{enumerate}
\item They saturate BPS bounds like $M=|Q|$ (extreme Reissner-Nordstr\"om
 solution).
\item Multicenter supersymmetric solutions are possible (Majumdar-Papapetrou
  multi-R-N-black hole solution) thanks to the equilibrium of forces
  $M_{i}M_{j}=Q_{i}Q_{j}$.
\item Their sources (possibly, branes) can be identified. 
\item They enjoy classical and quantum stability: results can be extrapolated
  to different domains (invariance under \textit{dualities}.).
\item They are more symmetric and have simpler functional forms that depend on
  a smaller number of independent functions.
\item They are easier to find: the off-shell equations of motion of
  supersymmetric configurations are related by the \textit{Killing Spinor
    Identities (KSIs)} \cite{Kallosh:1993wx}: if we denote the
  (l.h.s.~of the) bosonic equations of motion by $\mathcal{E}(\phi^{b}) \equiv
  \left. {\displaystyle\frac{\delta S}{\delta \phi^{b}}} \right|_{\phi^{f}=0}$
for a supersymmetric field configuration with Killing spinor $\epsilon$,
$\left.  \delta_{\epsilon}\phi^{f} \right|_{\phi^{f}=0}$ they are
\begin{equation}
\label{eq:ksis}
\mathcal{E}(\phi^{b})
\left. (\delta_{\epsilon}
\phi^{b})_{,f_{1}}
\right|_{\phi^{f}=0}
=0\, .
\end{equation}
These relations between the off-shell bosonic equations of motion
$\mathcal{E}(\phi^{b})$ are necessary conditions for supersymmetry.  We only
need to check a few equations of motion on a supersymmetric configuration.
The KSIs also constrain the possible sources enforcing cosmic censorship if we
require them to hold everywhere in spacetime \cite{Bellorin:2006xr}.  Finally,
they provide powerful consistency checks when we try to find large families of
supersymmetric solutions, as we are going to do.
\item In supersymmetric black-hole solutions there is an \textit{attractor
    mechanism} which suppresses \textit{primary scalar hair} and hints at a
  microscopic interpretation of the entropy \cite{Ferrara:1995ih}: consider a
  supersymmetric, static, spherically symmetric, asymptotically flat,
  black-hole solution given by the fields
\begin{equation}
  \{g_{rr}(r),F^{\Lambda}_{tr}(r),\star F^{\Lambda}_{tr}(r), \phi^{i}(r)\}\, .
\end{equation}
These solutions are fully characterized by the electric and magnetic charges
$q_{\Lambda}, p^{\Lambda}$ and the asymptotic values of the scalars
$\phi^{i}_{\infty}$. Supersymmetry imposes the saturation of the BPS bound:
$M= f(q_{\Lambda},p^{\Lambda},\phi^{i}_{\infty})$ for some function $f$.
It can be shown that \textit{at the event horizon} $r = r_{H}$ the scalars
$\phi^{i}$ and the metric function $r^{2}g_{rr}$ take their \textit{attractor}
value which  depends on the conserved charges $q_{\Lambda}, p^{\Lambda}$
and not on $\phi^{i}_{\infty})$:

\begin{equation}
\phi^{i}(r_{H}) =\phi^{i}_{attract}(q,p)\, ,   
\hspace{1cm}
r_{H}^{2}g_{rr}(r_{H})
=
4\pi S(q, p)\, .
\end{equation}

This proves that, at least for these supersymmetric black holes, the
Bekenstein-Hawking entropy $S(q, p)$ only depends on charges which will be
quantized, and therefore it is just a function of integers amenable to a
microscopic interpretation.
\end{enumerate}


\section{The search for all 4-d supersymmetric solutions} 

In his pioneering work \cite{Tod:1983pm} Tod showed that it is possible find
all the BPS solutions of pure $N=2,d=4$ supergravity (i.e.~solutions of the
Einstein-Maxwell theory). His result uses the doublet of spinors (we are in
$N=2$) as a basis in the Newman-Penrose formalism and has been generalized
(using the spinor-bilinear method developed in
Ref.~\cite{Gauntlett:2002fz}\footnote{See the references in
  \cite{Meessen:2010fh} for other methods and results in dimensions other than
  4.}) to include the coupling of more scalar and vector fields
\cite{Meessen:2006tu}, cosmological constant \cite{Caldarelli:2003pb} and
non-Abelian symmetries \cite{Huebscher:2007hj}. The supersymmetric solutions
found include regular black holes with non-Abelian fields, not in numerical
form as those in Refs.~\cite{Bartnik:1988am}, but in completely analytic form.

For $N>2$ there are further spinors containing information that cannot be
extracted with the Newman-Penrose formalism.  In \cite{Meessen:2010fh} it was
shown how to overcome those problems and determine the form of all the
timelike supersymmetric solutions\footnote{The $N\geq 2$ supersymmetric
  solutions fall in two cases: \textit{null} and \textit{timelike}.} of all $d=4$
ungauged supergravities using the spinor-bilinear method. Since they turn out
to be related to those of $N=2$ theories \cite{Huebscher:2007hj}, we briefly
review them first.


\subsection{The N=2 case}

The $N=2$ \textit{supergravity multiplet} is
\begin{equation}
\left\{ e^{a}{}_{\mu}, \psi_{I\, \mu}, A^{IJ}{}_{\mu}
\right\}\, ,\,\,\,\,
I,J,\dots=1,2\, ,\,\,\,\, \Rightarrow  A^{IJ}{}_{\mu}
=A^{0}{}_{\mu} \varepsilon^{IJ}\, .
\end{equation}
We can couple to it $n$ \textit{vector multiplets}
\begin{displaymath}
\left\{A^{i}{}_{\mu},\lambda^{i}{}_{I},Z^{i} \right\}\, ,\,\,\,\,\,
i=1,\cdots,n\, ,\,\,\,\, \Rightarrow A^{\Lambda}{}_{\mu}\, ,\,\,\,\,\,
  \Lambda=0,\cdots,n\, ,
\end{displaymath}
where the $Z^{i}$s are complex scalars, and $m$ \textit{hypermultiplets}
\begin{equation}
\left\{\zeta_{\alpha},q^{u}\right\}\, ,\,\,\,\,\,
u=1,\cdots,4m\, ,\,\,\,\,
\alpha=1,\cdots, 2m\, .
\end{equation}
The $n$ $Z^{i}$s are encoded into the $2\bar{n}$-dimensional symplectic
section ($\bar{n}=1+n$)
\begin{equation}
\mathcal{V}
=
\left( 
  \begin{array}{c}
\mathcal{L}^{\Lambda} \\ {\cal M}_{\Lambda} \\
  \end{array}
\right)\, ,
\hspace{1cm}
\langle \mathcal{V}\mid\mathcal{V}^{*}\rangle 
 =    
-2i\, . 
\end{equation}
This description of the scalars is extremely redundant but useful.

The action for the bosonic fields is
\begin{equation}
\begin{array}{rcl}
S 
& = & 
{\displaystyle\int} d^{4}x\sqrt{|g|}
\left[
R
+2\mathcal{G}_{ij^{*}}\partial_{\mu}Z^{i}\partial^{\mu}Z^{*j^{*}}
+2\mathsf{H}_{uv}\partial_{\mu}q^{u}\partial^{\mu}q^{v}
\right.
\\
& &
\hspace{2cm}
\left.
+2\Im{\rm m}\mathcal{N}_{\Lambda\Sigma} 
F^{\Lambda\, \mu\nu}F^{\Sigma}{}_{\mu\nu}
-2\Re{\rm e}\mathcal{N}_{\Lambda\Sigma} 
F^{\Lambda\, \mu\nu}\star F^{\Sigma}{}_{\mu\nu}
\right]\, ,
\end{array}
\end{equation}
where $\mathcal{G}_{ij^{*}}$ is the K\"ahler metric parametrized by the
$Z^{i}$s etc.\footnote{See Ref.~\cite{kn:n2d4} for a complete description of
  all the objects that appear in this action.}

The $N=2,d=4$ KSEs take the form
\begin{equation}
  \begin{array}{rcl}
\delta_{\epsilon}\psi_{I\, \mu} 
=  
\mathfrak{D}_{\mu}{\epsilon}_{I}
+\varepsilon_{IJ}\ T^{+}{}_{\mu\nu}\gamma^{\nu}\ \epsilon^{J}
& = & 0\, ,
\\
\delta_{\epsilon}\lambda^{iI} 
=
i\not\!\partial Z^{i}\epsilon^{I} \ 
+\ \varepsilon^{IJ}\not\! G^{i\, +}\ \epsilon_{J}
& = & 0\, ,
\\
\delta_{\epsilon}\zeta_{\alpha} 
= 
-i\mathbb{C}_{\alpha\beta}\ \mathsf{U}^{\beta I}{}_{u}\ \varepsilon_{IJ}\ 
\not\!\partial q^{u}\ {\epsilon}^{J}
& = & 0\, , 
\end{array}
\end{equation}
where ($\langle\, \cdot \mid \cdot \, \rangle$ is the symplectic product)
\begin{equation}
T^{+} = 
\langle\, \mathcal{V} \mid {\cal F}^{+} \, \rangle\, ,
\hspace{.5cm}
G^{i\, +} = \tfrac{i}{2}\mathcal{G}^{ij^{*}}
\langle\, \mathcal{D}_{j^{*}}\mathcal{V}^{*} \mid 
{\cal F}^{+} \, \rangle\, ,
\hspace{.5cm}
{\cal  F}^{+} \equiv 
\left(
  \begin{array}{c}
F^{\Lambda\, +} \\
\mathcal{N}^{*}_{\Lambda\Sigma} F^{\Sigma\, +} \\
\end{array}
\right)\, ,
\end{equation}
and  
$\mathfrak{D}$ is the Lorentz-K\"ahler-$SU(2)$-covariant derivative
($U(1)_{\rm Kahler}+SU(2) = U(2)$)
\begin{equation}
\mathfrak{D}_{\mu} \epsilon_{I} = 
(\partial_{\mu} +\tfrac{1}{4}  \omega_{\mu}{}^{ab}\gamma_{ab}\ 
+\ {\textstyle\frac{i}{2}}\ \mathcal{Q}_{\mu})\ 
\epsilon_{I} 
\ +\ \mathsf{A}_{\mu\, I}{}^{J}\ \epsilon_{J}\, ,
\end{equation}

\noindent
and where $\mathsf{U}^{\alpha I}{}_{u}(q)$ is the quaternionic-K\"ahler
$4m$-\textit{bein} \cite{kn:n2d4}.

Our goal is to find all the bosonic field configurations $\{e^{a}{}_{\mu},
A^{\Lambda}{}_{\mu}, Z^{i}, q^{u}\}$ such that the above KSEs admit at least
one solution $\epsilon^{I}(x)$. In the \textit{spinor-bilinear method}
\cite{Gauntlett:2002fz} we
\begin{enumerate}
\item Assume that one has a bosonic field configuration such that one solution
  $\epsilon^{I}$ exists.
\item Construct all the independent spinor bilinears with the commuting
  $\epsilon^{I}$ and find the equations they satisfy: (a) 
Due to the Fierz identities. (\textit{spinor-bilinear algebra}) and (b) Due to the KSEs.
\item Find their integrability conditions and show that they are also
  sufficient to solve the KSEs. At this point all supersymmetric
  configurations are determined.
\item Determine with the KSIs which equations of motion are independent for
  supersymmetric configurations and impose them on the supersymmetric
  configurations we just identified.
\end{enumerate}

The independent bilinears that we can construct with the $U(2)$ doublet of
Weyl spinors $\epsilon_{I}$ $I=1,2$ are:
\begin{enumerate}
\item A complex antisymmetric matrix of scalars $M_{IJ}\equiv
  \bar{\epsilon}_{I}\epsilon_{J}= X\varepsilon_{IJ}$. $X$ is an $SU(2)$
  singlet but has $U(1)$ K\"ahler weight.
\item A Hermitean matrix of vectors $V^{I}{}_{J\, a}\equiv
  i\bar{\epsilon}^{I}\gamma_{a}\epsilon_{J}$.
\end{enumerate}

The 4-d Fierz identities imply that $V_{a}\equiv V^{I}{}_{I\, a}$ is always
non-spacelike:

 \begin{equation}
V^{2}=-V^{I}{}_{J}\cdot V^{J}{}_{I} = 2
M^{IJ} M_{IJ}= 4|X|^{2}\geq 0\, .
\end{equation}

We only consider the timelike case $X\neq 0$ in which all $V^{I}{}_{J\, a}$
are independent. With them, ${\sigma}^{0}\equiv 1$ and the Pauli matrices
${\sigma}^{m}$ one can construct an orthonormal tetrad

\begin{equation}
V^{a}{}_{\mu}\equiv  {\textstyle\frac{1}{\sqrt{2}}}
V^{I}{}_{J\, \mu}({\sigma}^{a})^{J}{}_{I}\, ,
\hspace{1cm}
V^{I}{}_{J\, \mu}
=
{\textstyle\frac{1}{\sqrt{2}}}
V^{a}{}_{\mu}({\sigma}^{a})^{I}{}_{J}\, ,
\end{equation}

\noindent
in which $V^{0}=\sqrt{2}{V}$ is timelike and the $V^{m}$s are spacelike.

Observe that this construction does not work for $N >2$ where we have $U(N)$
vectors of spinors and we can only select 2 of them at the expense of breaking
manifest $U(N)$ invariance.

Apart from these equations (part of the spinor-bilinear algebra) the bilinears
$X$ and $V^{I}{}_{J}$ satisfy a number of equations that follow from the
assumption that $\epsilon^{I}$ satisfies the KSEs (is a Killing spinor). They
can be found in Refs.~\cite{Huebscher:2007hj}.

If we denote the (l.h.s.~of the) Einstein, Maxwell (and Bianchi) and scalar
equations of motion by
$\{\mathcal{E}^{\mu\nu},\mathcal{E}^{\mu},\mathcal{E}^{i},\mathcal{E}_{u}\}$
resp., then the KSIs of this theory imply:
\begin{enumerate}
\item $\mathcal{E}^{0m}=\mathcal{E}^{mn}=0$.
\item $\mathcal{E}^{m}=0$.
\item $\mathcal{E}_{u}=0$. This implies absence of attractor mechanism for the
  hyperscalars $q^{u}$.
\item $\mathcal{E}^{00} = -4|X| \langle\, \mathcal{E}^{0} \mid \, \Re{\rm
    e}(\mathcal{V}/{X}) \, \rangle$. This is related to the BPS bound. 
\item $0= \langle\, \mathcal{E}^{0} \mid \, \Im{\rm m}(\mathcal{V}/{X})
  \, \rangle$. This implies the absence of sources of NUT charge \cite{Bellorin:2006xr}.
\item $\mathcal{E}_{i^{*}} =
  2{\displaystyle\left(\frac{X}{X^{*}}\right)^{1/2}} \langle\, \mathcal{E}^{0}
  \mid \, \mathcal{D}_{i^{*}}\mathcal{V}^{*} \, \rangle$. This implies the
  existence of an attractor mechanism for the complex scalars $Z^{i}$.
\end{enumerate}
The only independent equations of motion that have to be imposed on $N=2,d=4$
supersymmetric configurations are, therefore, the zeroth components of the
Maxwell equations and Bianchi identities: $\mathcal{E}^{0}=0$.

The result can be summarized in the following recipe: all the supersymmetric
solutions of a $N=2,d=4$ supergravity can be constructed as follows:

\noindent
\textbf{{1}.} Define the $U(1)$-neutral real symplectic vectors $\mathcal{R}$
and $\mathcal{I}$
\begin{equation}
\mathcal{R}+i\mathcal{I} \equiv \mathcal{V}/{X}\, .
\end{equation}
No K\"ahler nor $SU(2)$ gauge-fixing are necessary.

\noindent
\textbf{{2}.} The components of $\mathcal{I}$ are given by a symplectic
vector real functions $\mathcal{H}$ harmonic in the 3-dimensional transverse
space with metric $\gamma_{\underline{m}\underline{n}}$:
\begin{equation}
\label{eq:harmonic}
\nabla^{2}_{(3)}\mathcal{H}=0\, .
\end{equation}
\textbf{{3}.} $\mathcal{R}$ can be found from $\mathcal{I}$ by using the
redundancy of the description provided by $\mathcal{V}$ which implies the
existence of relations between $\mathcal{R}$s and $\mathcal{I}$s known as
\textit{stabilization equations} and which may be very difficult to solve in
practice.

\noindent
\textbf{{4}.} The scalars $Z^{i}$ are given  by  the quotients
\begin{equation}
Z^{i}
=
\frac{\mathcal{V}^{i}/{X}}{\mathcal{V}^{0}/{X}}
=
\frac{\mathcal{R}^{i} +i\mathcal{I}^{i}}{\mathcal{R}^{0} 
+i\mathcal{I}^{0}}
\, .
\end{equation}
\textbf{{5}.}  The hyperscalars $q^{u}(x)$ are given by any mapping satisfying
\begin{equation}
\mathsf{ U}^{\alpha J}{}_{m}\ ({\sigma}^{m})_{J}{}^{I} \; =\; 0\, ,
\hspace{1cm}
\mathsf{U}^{\alpha J}{}_{n}\ \equiv\ V_{n}{}^{\underline{m}}
\partial_{\underline{m}}q^{u}\
\mathsf{U}^{\alpha J}{}_{u}\, .
\end{equation}
\textbf{{6}.} The metric takes the form
\begin{equation}
ds^{2} \; =\; 2|X|^{2} (dt+ \omega)^{2}   -\frac{1}{2|X|^{2}}
\gamma_{\underline{m}\underline{n}}dx^{m}dx^{n}\, .
\end{equation}
where 
\begin{equation}
|X|^{-2}
 =   
2\langle\,\mathcal{R}\mid \mathcal{I}\, \rangle\, ,
\hspace{1cm}
(d \omega)_{mn} 
= 
2\epsilon_{mnp}
\langle\,\mathcal{I}\mid \partial^{p}\mathcal{I}\, \rangle\, .
\end{equation}
$\gamma_{\underline{m}\underline{n}}$ is determined indirectly from the
hyperscalars: its spin connection $\varpi^{mn}$ in the basis
$\{{V}^{m}\}$ is related to the pullback of the $SU(2)$ connection of the
hyper-K\"ahler manifold $\mathsf{A}^{I}{}_{J}{}_{\mu}
=\frac{1}{\sqrt{2}}\mathsf{A}^{m}{}_{u} ({\sigma}^{m})^{I}{}_{J}\partial_{\mu}q^{u}$, by
\begin{equation}
\varpi_{m}{}^{np} = 
\varepsilon^{npq}
\mathsf{A}^{q}{}_{m}\, .
\end{equation}
\textbf{{7}.} The vector field strengths are 
\begin{equation}
\mathcal{F}=  -{\textstyle\frac{1}{2}} d (\mathcal{R}\hat{V})   
-{\textstyle\frac{1}{2}}\star(\hat{V}\wedge 
d\mathcal{I}) 
\, ,  
\hspace{1.5cm}
\hat{V} = 2\sqrt{2}|X|^{2}(dt+ \omega)\, . 
\end{equation}


\section{The all-$N$ case}

Dealing with all the $N>1,d=4$ supergravities simultaneously is possible
thanks to the formalism developed in Ref.~\cite{Andrianopoli:1996ve} which
generalizes that of the $N=2,d=4$ theories. It turns out that all 4-d
supergravity multiplets can be written in the form
\begin{equation}
\left\{ e^{a}{}_{\mu},\psi_{I\, \mu},A^{IJ}{}_{\mu},\chi_{IJK},
P_{IJKL\, \mu},\chi^{IJKLM}\right\}\, ,\,\,\,\,
I,J,\dots=1,\cdots, N\, ,
\end{equation}
and all vector multiplets can be written in the form 
\begin{equation}
\left\{A_{i\,    \mu},\lambda_{iI},{P}_{iIJ\, \mu},{\lambda}_{i}{}^{IJK}
\right\}\, ,\,\,\,\,\,
i=1,\cdots,n\, ,
\end{equation}
where $P_{IJKL\, \mu},{P}_{iIJ\, \mu}$ are the pullbacks of the scalar
Vielbeins. The price to pay for using this representation is that there is
some redundancy: all the fields that can be related by $SU(N)$ duality
relations, are:
\begin{itemize}
\item $N=4$: $P^{*\, i\, IJ} =  
\frac{1}{2}\varepsilon^{IJKL}P_{i\, KL}$,\,\,\, and\,\,\,
$\lambda_{iI}
=\frac{1}{3!}\varepsilon_{IJKL}
\lambda_{i}{}^{IJK}$.
\item $N=6$: $P^{*\, IJ} = 
\frac{1}{4!}\varepsilon^{IJK_{1}\cdots K_{4}}
P_{K_{1}\cdots K_{4}}$,\,\,\,
$\chi_{IJK} = \frac{1}{3!}\varepsilon_{IJKLMN}
\lambda^{IJK}$,\,\,\, 
$\chi^{I_{1}\cdots I_{5}} = 
\varepsilon^{I_{1}\cdots I_{5}J}\lambda_{J}$.
\item $N=8$: $P^{*\, I_{1}\cdots I_{4}} = 
\frac{1}{4!}\varepsilon^{I_{1}\cdots I_{4}J_{1}\cdots J_{4}}
P_{J_{1}\cdots J_{4}}$,\,\,\,
$\chi_{I_{1}I_{2}I_{3}} =
\frac{1}{5!}\varepsilon_{I_{1}I_{2}I_{3}J_{1}\cdots J_{5}}
\chi^{J_{1}\cdots J_{5}}$.
\end{itemize}
All these constraints must be carefully taken into account.

The scalars are encoded into the $2\bar{n}$-dimensional ($\bar{n}\equiv
n+\frac{N (N -1)}{2}$) symplectic vectors 
\begin{equation}
\mathcal{V}_{IJ}
=
\left( 
  \begin{array}{c}
f^{\Lambda}{}_{IJ} \\ h_{\Lambda\, IJ} \\
  \end{array}
\right)\, ,
\hspace{1cm}
\mathrm{and}
\hspace{1cm}
\mathcal{V}_{i}
=
\left( 
  \begin{array}{c}
   f^{\Lambda}{}_{i} \\ h_{\Lambda\, i} \\
  \end{array}
\right)\, ,
\hspace{1cm}
\Lambda=1,\cdots, \bar{n}\, ,
\end{equation}
normalized 
\begin{equation}
\langle \mathcal{V}_{IJ}\mid\mathcal{V}^{*\, KL}\rangle 
 =    
-2i\delta^{KL}{}_{IJ}\, , 
\hspace{1cm}
\langle \mathcal{V}_{i}\mid\mathcal{V}^{*\, j}\rangle 
 =    
-i\delta_{i}{}^{j}\, . 
\end{equation}
They can be combined into the  $Usp(\bar{n},\bar{n})$ matrix
\begin{equation}
U \equiv 
{\textstyle\frac{1}{\sqrt{2}}}
\left(
  \begin{array}{cc}
f+ih 
& 
f^{*}+ih^{*} 
\\
f-ih 
& 
f^{*}-ih^{*} 
\\
  \end{array}
\right)\, .  
\end{equation}
They reduce to the standard  sections of the $N=2$ case
\begin{equation}
\mathcal{V}_{IJ}= \mathcal{V}\varepsilon_{IJ}\, ,
=
\left( 
\begin{array}{c}
{\mathcal{L}}^{\Lambda}\varepsilon_{IJ}\\ 
{\mathcal{M}}_{\Lambda}\varepsilon_{IJ} \\
  \end{array}
\right)\, ,
\hspace{1cm}
\mathrm{and}
\hspace{1cm}
 \mathcal{V}_{i}= \mathcal{D}_{i}\mathcal{V}
=\left( 
\begin{array}{c}
f^{\Lambda}{}_{i}\\ 
h_{\Lambda\, i} \\
  \end{array}
\right)\, .
\end{equation}
The graviphotons $A^{IJ}{}_{\mu}$ do not appear directly, only through the
``dressed'' vectors
\begin{equation} 
A^{\Lambda}{}_{\mu}\equiv 
{\textstyle\frac{1}{2}}f^{\Lambda}{}_{IJ}
A^{IJ}{}_{\mu}
+f^{\Lambda}{}_{i} A^{i}{}_{\mu}\, .
\end{equation}
The action for the bosonic fields is
\begin{equation}
\begin{array}{rcl}
S 
& = & 
{\displaystyle\int} d^{4}x\sqrt{|g|}
\left[
R
+2\Im{\rm m}\mathcal{N}_{\Lambda\Sigma} 
F^{\Lambda\, \mu\nu}F^{\Sigma}{}_{\mu\nu}
-2\Re{\rm e}\mathcal{N}_{\Lambda\Sigma} 
F^{\Lambda\, \mu\nu}\star F^{\Sigma}{}_{\mu\nu}
\right.
\\
& &
\hspace{2cm}
\left.
+\frac{2}{4!}\alpha_{1}P^{*\, IJKL}{}_{\mu}P_{IJKL}{}^{\mu}
+\alpha_{2}P^{*\, iIJ}{}_{\mu}P_{iIJ}{}^{\mu}
\right]\, ,
\end{array}
\end{equation}
where 
\begin{equation}
\mathcal{N}=hf^{-1}=\mathcal{N}^{T}\, ,
\hspace{1cm}
h_{\Lambda} = 
\mathcal{N}_{\Lambda\Sigma}f^{\Sigma}\, .
\hspace{1cm}
\mathfrak{D}h_{\Lambda}=
\mathcal{N}^{*}_{\Lambda\Sigma}\mathfrak{D}f^{\Lambda}\, .
\end{equation}
The $N$-specific constraints must be taken into account to find the
e.o.m.:
\begin{description}
\item[For $N=2$:]
$\mathcal{E}^{iIJ}  =  \mathfrak{D}^{\mu} P^{*\, iIJ}{}_{\mu}
+2T^{i\, -}{}_{\mu\nu} T^{IJ\, -\, \mu\nu}
+P^{*\, iIJ\, A} P^{*\, jk}{}_{A}T_{j}{}^{+}{}_{\mu\nu}T_{k}{}^{+\, \mu\nu}$.

\item[For $N=3$:] 
$\mathcal{E}^{iIJ}  =  \mathfrak{D}^{\mu} P^{*\, iIJ}{}_{\mu}  
+2T^{i\, -}{}_{\mu\nu}T^{IJ\, -\, \mu\nu}$.

\item[For $N=4$:]
$\left\{
\begin{array}{rcl}
\mathcal{E}^{IJKL} 
& = & 
\mathfrak{D}^{\mu}P^{*\, IJKL}{}_{\mu}
+6 T^{[IJ|-}{}_{\mu\nu} T^{|KL]-\, \mu\nu}
\\
& & \\
& & 
+P^{*\, IJKL\, A} P^{*\, ij}{}_{A}T_{i}{}^{+}{}_{\mu\nu}
T_{j}{}^{+\, \mu\nu}\, ,
\\
& & \\
\mathcal{E}^{iIJ} 
& = & 
\mathfrak{D}^{\mu}P^{*\, iIJ}{}_{\mu}
+T^{i\, -}{}_{\mu\nu}T^{IJ\, -\, \mu\nu}
+\frac{1}{2}\varepsilon^{IJKL}T_{i}{}^{+}{}_{\mu\nu}T_{KL}{}^{+\, \mu\nu}\, ,
\end{array}
\right.
$

etc.
\end{description}
The supersymmetry transformations of the fermionic fields are
\begin{equation}
  \begin{array}{rcl}
\delta_{\epsilon} \psi_{I\mu} 
& = &
\mathfrak{D}_{\mu}\epsilon_{I}
+T_{IJ}{}^{+}{}_{\mu\nu}\gamma^{\nu}
\epsilon^{J}\, ,
\\ 
\delta_{\epsilon} \chi_{IJK} 
& = & 
-\frac{3i}{2} \not\! T_{[IJ}{}^{+}
\epsilon_{K]}
+i\not\! P_{IJKL}\epsilon^{L}\, ,
\\
\delta_{\epsilon}\lambda_{iI}  
& = &  
-\frac{i}{2} \not\! T_{i}{}^{+}
\epsilon_{I}
+i\not\! P_{iIJ}\epsilon^{J}\, ,
\\
\delta_{\epsilon} \chi_{IJKLM} 
& = &
-5i \not\! P_{[IJKL}\epsilon_{M]}
+\frac{i}{2}\varepsilon_{IJKLMN} \not\! T^{-}
\epsilon^{N}
+\frac{i}{4}\varepsilon_{IJKLMNOP} \not\! T^{NO-}
\epsilon^{P}\, ,
\\
\delta_{\epsilon} \lambda_{iIJK} 
& = &
-3i \not\! P_{i[IJ}\epsilon_{K]}
+\frac{i}{2} \varepsilon_{IJKL} \not\! T_{i}{}^{-}
\epsilon^{L}
+\frac{i}{4}\varepsilon_{IJKLMN}\not\! T^{LM-}
\epsilon_{N}\, ,
\end{array}
\end{equation}
where  the graviphoton and matter vector field strengths are
\begin{equation} 
T_{IJ}{}^{+}
= 
\langle\, \mathcal{V}_{IJ} \mid {\cal F}^{+} \, \rangle\, ,
\,\,\,\,\,\,\,
T_{i}{}^{+}
=
\langle\, \mathcal{V}_{i} \mid {\cal  F}^{+} \, \rangle\, ,
\,\,\,\,\,\,\,\,
{\cal  F}_{\Lambda}{}^{+}= \mathcal{N}^{*}_{\Lambda\Sigma}{F}^{\Sigma\, +}\, ,
\end{equation}
and where 
\begin{equation}
\mathfrak{D}_{\mu}\epsilon_{I} \equiv 
\nabla_{\mu}\epsilon_{I}
-\epsilon_{J}\Omega_{\mu}{}^{J}{}_{I}\, ,  
\end{equation}
and $\Omega_{\mu}{}^{J}{}_{I}$ is the pullback of the
connection of the  scalar manifold ($\subset U(N)$).

For all values of $N$ the independent KSEs take the form
\begin{equation}
\begin{array}{rcl}
\mathfrak{D}_{\mu}\epsilon_{I}
+T_{IJ}{}^{+}{}_{\mu\nu}\gamma^{\nu}
\epsilon^{J} 
& = & 0\, ,
\\ 
\not\!P_{IJKL}\epsilon^{L}
-{\textstyle\frac{3}{2}} \not \!T_{[IJ}{}^{+}
\epsilon_{K]}
& = & 
0\, ,
\\
\not\!P_{i\, IJ}\epsilon^{J}
-{\textstyle\frac{1}{2}} \not \!T_{i}{}^{+}
\epsilon_{I}
& = & 
0\, ,
\\
\not\!P_{[IJKL}\epsilon_{M]}
& = & 
0\, ,
\\
\not\!P_{i\, [IJ}\epsilon_{K]}
& = & 
0\, .
\\
\end{array}
\end{equation}
The last two KSEs should only be considered for $N=5$  and $N=3$,
resp.

Again, our goal is to find all the bosonic field configurations
$\{e^{a}{}_{\mu}, A^{\Lambda}{}_{\mu}, P_{IJKL\, \mu}, P_{i\, IJ\, \mu}\}$
such that the above KSEs admit at least one solution $\epsilon^{I}$ following
the spinor-bilinear method. We only consider the timelike case.   

First, we construct all the possible independent
bilinears with one $U(N)$ vector of Weyl spinors $\epsilon_{I}$ and compute
the spinor-bilinears algebra.  The independent bilinears are:
\begin{enumerate}
\item A complex antisymmetric matrix of scalars $M_{IJ}\equiv
  \bar{\epsilon}_{I}\epsilon_{J}=-{M}_{JI}$. 
\item A Hermitean matrix of vectors $V^{I}{}_{J\, a}\equiv
  i\bar{\epsilon}^{I}\gamma_{a}\epsilon_{J}$.
\end{enumerate}
The Fierz identities imply the following properties for them:
\begin{enumerate}
\item $M_{I[J}M_{KL]}=0$, so $\mathrm{rank}\, (M_{IJ}) \leq 2$.
\item $V_{a}\equiv V^{I}{}_{I\, a}$ is always non-spacelike: $V^{2}= 2 M^{IJ}
  M_{IJ}\equiv 2|M|^{2}\geq 0$.
\item We can choose a tetrad $\{e^{a}{}_{\mu}\}$ such that
  $e^{0}{}_{\mu}\equiv \frac{1}{\sqrt{2}}|M|^{-1}{V}_{\mu}$. Then,
  \textit{defining} $V^{m}{}_{\mu} \equiv |M| e^{m}{}_{\mu}$ we can decompose
\begin{equation}
V^{I}{}_{J\, \mu} = \tfrac{1}{2}\mathcal{J}^{I}{}_{J}V_{\mu}
+\tfrac{1}{\sqrt{2}}(\sigma^{m})^{I}{}_{J}
V^{m}{}_{\mu}\, ,
\end{equation}
where $\mathcal{J}^{I}{}_{J} = 2 M^{IK}
M_{JK}|M|^{-2}$ is a rank 2 projector (Tod):
\begin{equation}
\mathcal{J}^{2} = \mathcal{J}\, ,
\hspace{1cm}
\mathcal{J}^{I}{}_{I}=+2\, ,
\hspace{1cm}
\mathcal{J}^{I}{}_{J}\epsilon^{J}
=\epsilon^{I} \, .
\end{equation}
\end{enumerate}

The main properties satisfied by the three $\sigma^{m}$
matrices are:
\begin{equation}
  \begin{array}{rcl}
\sigma^{m}\sigma^{n} 
& = & 
\delta^{mn}\mathcal{J} 
+i\varepsilon^{mnp}\sigma^{p}\, ,
\\
\mathcal{J}\sigma^{m} 
&  = &  
\sigma^{m} \mathcal{J} 
= \sigma^{m}\, ,\\
(\sigma^{m})^{I}{}_{I} 
& = & 0\, ,
\\  
\mathcal{J}^{K}{}_{J}\mathcal{J}^{L}{}_{I}
& = & 
\frac{1}{2}\mathcal{J}^{K}{}_{I}\mathcal{J}^{L}{}_{J}
+
\frac{1}{2}(\sigma^{m})^{K}{}_{I}
(\sigma^{m})^{L}{}_{J}\,
,
\\
M_{K[I}(\sigma^{m})^{K}{}_{J]} 
& = & 
0\, ,\\  
2|M|^{-2}M_{LI}
(\sigma^{m})^{I}{}_{J}M^{JK} 
& = & 
(\sigma^{m})^{K}{}_{L}\, ,
\\  
\end{array}
\end{equation}
Summarizing: $\{\mathcal{J},\sigma^{1}, \sigma^{2},\sigma^{3}\}$ is an
$x$-dependent basis of a $\mathfrak{u}(2)$ subalgebra of $\mathfrak{u}(N)$ in
the 2-dimensional eigenspace of $\mathcal{J}$ of eigenvalue $+1$ and provide a
basis in the space of Hermitean matrices $A$ satisfying $\mathcal{J}
A\mathcal{J}=A$

These bilinears also satisfy a number of equations that follow from the
assumption that $\epsilon^{I}$ is a Killing spinor. They can be found in
Ref.~\cite{Meessen:2010fh}. Also from this assumption follow the KSIs
that constrain the (l.h.s.~of the) Einstein equations, Maxwell equations and
Bianchi identities and scalar equations of motion, denoted by
$\{\mathcal{E}^{\mu\nu},\mathcal{E}^{\mu},\mathcal{E}^{IJKL},\mathcal{E}^{i\,
  IJ}\}$ resp. Defining $\tilde{\mathcal{J}}{}^{I}{}_{J} \equiv
\delta^{I}{}_{J}-\mathcal{J}{}^{I}{}_{J}$ we get
\begin{description}
\item[1.] $\mathcal{E}^{0m}=\mathcal{E}^{mn}=0$.
\item[2.] $\mathcal{E}^{m}=0$.
\item[3.] $
\left\{
  \begin{array}{rcl}
\mathcal{E}^{MNPQ}
\mathcal{J}^{[I}{}_M \tilde{\mathcal{J}}{}^{J}{}_{N}
\tilde{\mathcal{J}}{}^{K}{}_{P} \tilde{\mathcal{J}}{}^{L]}{}_{Q}
& = & 0\, ,\\
& & \\
\mathcal{E}^{i\, MN}\mathcal{J}^{[I}{}_M 
\tilde{\mathcal{J}}{}^{J]}{}_{N}
& = & 0\, ,\\
\end{array}
\right.
$
(No attractor mechanism). 
\item[4.] $\mathcal{E}^{00} = -2\sqrt{2} \langle\, \mathcal{E}^{0} \mid \,
  \Re{\rm e}\, {\displaystyle\left(\mathcal{V}_{IJ}\frac{M^{IJ}}{|M|}\right)}
  \, \rangle$, (BPS bound)
\item[5.] $\langle\, \mathcal{E}^{0} \mid \, \Im{\rm m}\,  
{\displaystyle\left(\mathcal{V}_{IJ}\frac{M^{IJ}}{|M|}\right)}
  \, \rangle$, (No NUT charge).
\item[6.] $
\left\{
  \begin{array}{l}
\mathcal{E}^{MNPQ}
\mathcal{J}^{[I}{}_M \mathcal{J}{}^{J}{}_{N}
\tilde{\mathcal{J}}{}^{K}{}_{P} \tilde{\mathcal{ J}}{}^{L]}{}_{Q}\, ,
\\
\\
\mathcal{E}^{i\, MN}\mathcal{J}^{[I}{}_M \mathcal{J}{}^{J]}{}_{N}
\, ,\\
\end{array}
\right.  $ are related to $\mathcal{E}^{0}$ 
(attractor mechanism).

The precise form of the relation depends on $N$:
\begin{description}
\item[$N=3$:]\,\,\,\,\, $\mathcal{E}^{i\, IJ}
 =  
-2\sqrt{2}
{\displaystyle\frac{M^{IJ}}{|M|}}
\langle\,\mathcal{E}^{0} \mid \mathcal{V}^{*\, i}\, \rangle \, ,
$
\item[$N=4$:]\,\,\,\,\,
$
\left\{
\begin{array}{rcl}
\mathcal{E}^{IJKL} 
& = &  
-2\sqrt{2}
{\displaystyle\frac{M^{[IJ|}}{|M|}}
\langle\,\mathcal{E}^{0} \mid 
\mathcal{V}^{*\, |KL]}\, \rangle\, , 
\\
& & \\
\mathcal{E}_{iIJ}
& = &  
-2\sqrt{2}\left\{
{\displaystyle\frac{M_{IJ}}{|M|}}
\langle\,\mathcal{E}^{0} \mid \mathcal{V}_{i}\, \rangle    
+\frac{1}{2}\varepsilon_{IJKL}
{\displaystyle\frac{M^{KL}}{|M|}}
\langle\,\mathcal{E}^{0} \mid \mathcal{V}^{*\, i}\, \rangle    
\right\}\, ,
\\
\end{array}
\right.
$

etc.
\end{description}
\end{description}
The only independent equations of motion that have to be imposed on \textit{any}
${d=4}$ supersymmetric configuration are again $\mathcal{E}^{0}=0$. Observe
that there are scalars which play a r\^ole analogous to that of the complex
scalars in the $N=2$ theories and are subject to an attractor mechanism and
scalars which play a role analogous to the hyperscalars and are not subject to
any such mechanism. 

Analogously to the $N=2$ case, we find that the construction of any timelike
supersymmetric solution proceeds as follows:

\textbf{{I}.} Choose the $U(2)$ subgroup determining
the associated $N=2$ \textit{truncation}:

\begin{enumerate}
\item Choose $x$-dependent rank-2, $N \times N$ complex antisymmetric
  $M_{IJ}$.  With it we construct the projector $\mathcal{J}^{I}{}_{J}
  \equiv 2 |M|^{-2}M^{IK} M_{JK}$.
Supersymmetry requires that
\begin{equation}
  \mathfrak{D}\mathcal{J}\equiv d\mathcal{J} 
-[\mathcal{J}, \Omega ]=0\, ,  
\end{equation}
%
%
\item Choose three $N \times N$, Hermitean, traceless, $x$-dependent
  $(\sigma^{m})^{I}{}_{J}$, satisfying the same properties as the Pauli
  matrices in the subspace preserved by $\mathcal{J}$.

  We also have to impose the constraint
\begin{equation}
\mathcal{J}d\sigma^{m}\mathcal{J}=0\, .
\end{equation}
\end{enumerate}
Once the $U(2)$ subgroup has been chosen, we can split the Vielbeins
$P_{IJKL\, \mu}$ and $P_{i\, IJ\, \mu}$, into associated to the would-be
vector multiplets in the $N=2$ \textit{truncation}
\begin{equation} 
P_{IJKL}\, 
\mathcal{J}^{I}{}_{[M}\mathcal{J}^{J}{}_{N} \tilde{\mathcal{J}}{}^{K}{}_{P} 
\tilde{\mathcal{J}}{}^{L}{}_{Q]}\, , \,\,\,\,\,
  \mathrm{and}\,\,\,\,\,\, P_{i\, IJ}\, 
\mathcal{J}^{I}{}_{[K}\mathcal{J}^{J}{}_{L]}\, ,
\end{equation}
which are driven by the \textit{attractor mechanism} (\textit{i.e.}~they are
determined by the electric and magnetic charges) and those associated to the
\textit{hypermultiplets}
\begin{equation}
P_{IJKL}\, \mathcal{J}^{I}{}_{[M} \tilde{\mathcal{J}}{}^{J}{}_{N}
\tilde{\mathcal{J}}{}^{K}{}_{P} \tilde{\mathcal{J}}{}^{L}{}_{Q]}
\, ,\,\,\,\,\,
\mathrm{and}\,\,\,\,\,\,
P_{i\,  IJ}\, \mathcal{J}^{I}{}_{[K} 
\tilde{\mathcal{J}}{}^{J}{}_{L]}\, .
\end{equation}
which are not.

In \textit{hyper}-less solutions (\textit{e.g.} black holes) the $\sigma^{m}$s
matrices are not needed at all.

\textbf{ {II}.} After the choice of $U(2)$ subgroup,
the solutions are constructed:

\textbf{{1}.} Define the real symplectic vectors $\mathcal{R}$ and
$\mathcal{I}$
\begin{equation}
\mathcal{R}+i\mathcal{I} \equiv  |M|^{-2}\mathcal{V}_{IJ}M^{IJ}\, .
  \end{equation}
($U(N)$ singlets $\Rightarrow$ no $U(N)$ gauge-fixing necessary) 

\textbf{{2}.} The components of $\mathcal{I}$ are given by a symplectic vector
real functions $\mathcal{H}$ harmonic in the 3-dimensional transverse space
with metric $\gamma_{\underline{m}\underline{n}}$ Eq.~(\ref{eq:harmonic}).

\textbf{{3}.} $\mathcal{R}$ is to be be found from $\mathcal{I}$ exploiting
again the redundancy in the description of the scalars by the sections
$\mathcal{V}_{IJ}, \mathcal{V}_{i}$.

\textbf{{4}.} The metric is
\begin{equation}
ds^{2} \; =\; |M|^{2} (dt+ \omega)^{2} 
  -|M|^{-2} \gamma_{\underline{m}\underline{n}}dx^{m}dx^{n}\, .
\end{equation}
where  
\begin{equation}
|M|^{-2} 
=   
(M^{IJ}M_{IJ})^{-2}=
\langle\,\mathcal{R}\mid \mathcal{I}\, \rangle\, ,
\hspace{1cm}
(d \omega)_{mn} 
=  
2\epsilon_{mnp}
\langle\,\mathcal{I}\mid \partial^{p}\mathcal{I}\, \rangle\, .
\end{equation}
$\gamma_{\underline{m}\underline{n}}$ is determined indirectly from the
would-be \textit{hypers} in the associated $N=2$ \textit{truncation}
and its curvature vanishes when those scalars vanish.

Its spin connection $\varpi^{mn}$is related to $\Omega$, by
\begin{equation}
\varpi^{mn} = 
i\varepsilon^{mnp}\mathrm{Tr}\, [\sigma^{p}\Omega]\, .
\end{equation}
(Observe that only the $\mathfrak{su}(2)$ components of $\Omega$ contribute
to $\varpi^{mn}$.

\textbf{{5}.} The vector field strengths are 
\begin{equation}
F=  -{\textstyle\frac{1}{2}} d (\mathcal{R}\hat{V})   
-{\textstyle\frac{1}{2}}\star(\hat{V}\wedge d\mathcal{I}) \, ,  
\hspace{1.5cm}
\hat{V} = \sqrt{2}|M|^{2}(dt+ \omega)\, . 
\end{equation}

\textbf{{6}.} The scalars in the \textit{vector multiplets} in the associated
$N=2$ \textit{truncation}
\begin{equation}
P_{IJKL}\,  \mathcal{J}^{I}{}_{[M}\mathcal{J}^{J}{}_{N} 
\tilde{\mathcal{J}}{}^{K}{}_{P} \tilde{\mathcal{J}}{}^{L}{}_{Q]}\, ,
\,\,\,\,\,
\mathrm{and}\,\,\,\,\,\,
P_{i\, IJ}\,   \mathcal{J}^{I}{}_{[K}\mathcal{J}^{J}{}_{L]}\, , 
\end{equation}
can be found from $\mathcal{R}$ and $\mathcal{I}$, while those in
the \textit{hypers} must be found independently by solving
\begin{equation}
\begin{array}{rcl}
P_{IJKL\, m}\, \mathcal{J}^{I}{}_{[M}
\tilde{\mathcal{J}}{}^{J}{}_{N} \tilde{\mathcal{J}}{}^{K}{}_{P}
\tilde{\mathcal{J}}{}^{L}{}_{Q]} (\sigma^{m})^{Q}{}_{R} 
& = & 
0\, ,
\\
& & \\  
P_{i\, IJ\, m}\, \mathcal{J}^{I}{}_{[K}
\tilde{\mathcal{J}}{}^{J}{}_{L]}(\sigma^{m})^{L}{}_M 
& = & 
0\, ,
\\
\end{array}
\end{equation}
which solve their equations of motion according to the KSIs.


\section{Conclusions}

Supersymmetric solutions play an extremely important r\^ole in many recent
developments. We have determined the general form of all the timelike
supersymmetric solutions of all ungauged $d=4$ supergravities and we have
proven the relation between the timelike supersymmetric solutions of all $d=4$
supergravities and those of the $N=2$ theories (conjectured for black-hole
solutions in \cite{Ferrara:2006yb}. 

Much work remains to be done in order to make explicit the construction of the
solutions: one has to find general parametrizations of the matrices $M^{IJ}$
and ${\cal J}^{I}{}_{J}$, solve the \textit{stabilization equations}, impose the
covariant constancy of $\mathcal{J}$ etc. However, this result will allow us
to have in explicit, analytic way the most general U-duality invariant
families of $d=4$ black-hole solutions. Work in this direction is already in
progress.


\subsection*{Acknowledgments}

  This has been supported in part by the Spanish Ministry of Science and
  Education grant FPA2009-07692, the Comunidad de Madrid grant HEPHACOS
  S2009ESP-1473 and the Spanish Consolider-Ingenio 2010 program CPAN
  CSD2007-00042. The author would like to thank M.M.~Fern\'andez for her
  permanent support.


\end{document}